\documentclass{ptephy_v1}

\usepackage{graphicx}
\usepackage{color}

\renewcommand{\vec}[1]{\mbox{\boldmath $#1$}}

\begin{document}

\title{
Resonance width for a particle-core coupling model 
with a square-well potential} 

\author{
\name{K. Hagino}{1,2,3}, 
\name{H. Sagawa}{4,5},
\name{S. Kanaya}{6}, and 
\name{A. Odahara}{6}
}

\address{
\affil{1}{ 
$^1$Department of Physics, Tohoku University, Sendai 980-8578,  Japan} \\
\affil{2}{$^2$Research Center for Electron Photon Science, Tohoku University, 1-2-1 Mikamine, Sendai 982-0826, Japan} \\
\affil{3}{$^3$Department of Physics, Kyoto University, Kyoto 606-8502,  Japan} \\
\affil{4}{$^4$RIKEN Nishina Center, Wako 351-0198, Japan} \\
\affil{5}{$^5$Center for Mathematics and Physics,  University of Aizu, 
Aizu-Wakamatsu, Fukushima 965-8560,  Japan} \\
\affil{6}{$^6$Department of Physics, Osaka University, Osaka 560-0043, Japan}
}

\begin{abstract}%
We derive a compact formula for the width of 
a multi-channel resonance state. To this end, we use a deformed 
square-well potential and solve the coupled-channels equations. 
We obtain the $S$-matrix in the Breit-Wigner form, 
from which partial widths can be extracted. We apply the resultant formula 
to a deformed nucleus and discuss the behavior of partial width for 
an $s$-wave channel. 
\end{abstract}

\subjectindex{D13,D29}

\maketitle

\section{Introduction}

Much attention has been paid recently to the study of unbound states in nuclei near the neutron and proton drip lines, 
stimulated by the rapid progress of radioactive ion beam experiments \cite{Pfutzner12,WD97,Nakamura17}. 
This applies both to unbound nuclei beyond the drip lines 
and to 
states in bound nuclei above the threshold of a 
particle emission. 
A particular interest is in 
single-particle resonance states, which decay 
to a neighboring nucleus 
by emitting one nucleon. 
Such resonances also have a large impact on the r-process and rp-process nucleosyntheses. 
Bohr and Mottelson have derived a simple formula for the width 
of a single-particle resonance state using a spherical square-well 
potential \cite{bm1}. The formula has been used, e.g., 
in Ref. \cite{MDG94}. 

In this paper, we extend the formula derived by Bohr and Mottelson to 
multi-channel resonances. That is, we discuss 
resonances in a particle-core system, in which several angular momentum 
components are coupled together due to excitations of the core nucleus. 
Such channel coupling effects 
have been known to play an important role in one nucleon decays of 
unbound nuclei 
\cite{KBNV00,BKNV00,MF00,DE00,ED00,DE01,Hagino01,Karny03,Fossez16}
(see also Ref. \cite{WN18} for a two-proton decay). 
Single-particle resonances in a deformed potential have also been 
discussed, e.g., in Refs. \cite{MFL98,Hagino-Giai04,YH05,Hamamoto}. 
In particular, the coupling between $s$-wave and $d$-wave 
components often plays a crucial role in neutron-rich 
nuclei close to the drip-line \cite{YH05,Hamamoto}. 
The extended formula presented in this paper 
will provide a simple 
estimate of partial and total widths for such 
multi-channel resonance states. 

The paper is organized as follows. In Sec. II, we consider a spherical 
square-well potential and summarize the formula of Bohr and Mottelson. 
In Sec. III, we extend the formula 
to multi-channel cases. 
We apply the extended formula to a deformed nucleus by taking into account 
the rotational excitations of the core nucleus, which results in a mixture 
of $s$ and $d$ waves. We discuss the partial width of the $s$-wave channel, 
for which a resonance may not exist in the absence of the coupling to the 
$d$-wave component. 
We then summarize the paper in Sec. IV. 

\section{Spherical square-well potential}

In this paper, we consider a two-body system with a neutron and a core nucleus. 
Let us first consider a positive energy state of the system 
in a spherical 
square-well potential given by,
\begin{equation}
V(r)=V_0\,\theta(R-r), 
\label{square}
\end{equation}
where $V_0 (<0)$ and $R$ are the depth and the range of the potential, 
respectively. Here, $\theta(x)$ is the step function. 

We shall closely follow Ref. \cite{bm1} and derive 
the $S$-matrix for this potential. 
Writing the wave function for the relative motion as 
\begin{equation}
\Psi(\vec{r})=\frac{u_l(r)}{r}\,Y_{lm}(\hat{\vec{r}}),
\end{equation}
where $l$ and $m$ are the orbital angular momentum and its $z$ component, 
respectively, and $Y_{lm}(\hat{\vec{r}})$ is the spherical harmonics, 
the Schr\"odinger equation for the radial wave function, $u_l(r)$, 
reads, 
\begin{equation}
\left(-\frac{\hbar^2}{2\mu}\frac{d^2}{dr^2}+V(r)+\frac{l(l+1)\hbar^2}{2\mu r^2}
-E\right)u_l(r)=0, 
\end{equation}
where 
$\mu$ is the reduced mass and $E (>0)$ is the energy. 
The solution of this equation with the spherical square-well potential reads
\begin{eqnarray}
u_l(r)&=&A_l Krj_l(Kr)~~~~~~~~~~~~~~~~~~~~~~~~~~~(r < R), \label{bc0-3} \\
&=&krh_l^{(-)}(kr)-U_l(E)krh_l^{(+)}(kr)~~~~(r\geq R),
\label{bc0-2}
\end{eqnarray}
with $K=\sqrt{\frac{2\mu}{\hbar^2}(E-V_0)}$ and 
$k=\sqrt{\frac{2\mu}{\hbar^2}E}$. 
Here, $A_l$ is a constant and $U_l$ is the $S$-matrix, which are given in terms 
of the phase shift $\delta_l$ as $U_l=e^{2i\delta_l}$. 
$h_l^{(\pm)}(x)$ is the spherical Hankel functions, which are given 
by $h_l^{(\pm)}(x)=-n_l(x)\pm i j_l(x)$ using the spherical Bessel 
function, $j_l(x)$, and the spherical Neumann function, $n_l(x)$ \cite{phase}. 

From Eq. (\ref{bc0-2}), 
one obtains 
\begin{equation}
U_l=\frac{L_l-S_l+iP_l}{L_l-S_l-iP_l}\,e^{2i\phi_l}, 
\label{s-matrix0}
\end{equation}
where 
\begin{equation}
L_l\equiv R\left(\frac{d}{dr}u_l(r)\right)_{r=R}\,\frac{1}{u_l(R)}, 
\end{equation}
is the logarithmic derivative of the wave function at $r=R$. 
Notice that the logarithmic derivative of the wave function is 
continuous at $r=R$, and $L_l$ can be evaluated using either 
Eq. (\ref{bc0-3}) or Eq. (\ref{bc0-2}).  
$S_l$ and $P_l$ in Eq. (\ref{s-matrix0}) are defined as, 
\begin{eqnarray}
S_l&=&\frac{G_l(R)G_l'(R)+F_l(R)F_l'(R)}{G_l(R)^2+F_l(R)^2}, 
\label{delta}
\\
P_l&=&\frac{G_l(R)F_l'(R)-F_l(R)G_l'(R)}{G_l(R)^2+F_l(R)^2} 
=\frac{kR}{G_l(R)^2+F_l(R)^2}, 
\label{s}
\end{eqnarray}
respectively. 
Here, we have followed Ref. \cite{bm1} and introduced shorthanded 
notations defined as, 
\begin{eqnarray}
F_l(r)&\equiv&kr\,j_l(kr),~~~G_l(r)\equiv -kr\,n_l(kr), 
\label{FandG}
\\
F'_l(r)&\equiv&R\frac{dF_l}{dr},~~~~~~G'_l(r)\equiv R\frac{dG_l}{dr}. 
\label{FandG2}
\end{eqnarray} 
Note that $S_l$ in Eq. (\ref{delta}) and $P_l$ in Eq. (\ref{s}) 
correspond to the shift function and the penetrability in the R-matrix 
method, respectively \cite{Nunes,LT58,DB10,KN04}.
The phase factor $e^{2i\phi_l}$ in Eq. (\ref{s-matrix0})  
is the hard sphere scattering term (which is a part of the 
background $S$-matrix) given by 
\begin{equation}
e^{2i\phi_l}=\frac{G_l(R)-iF_l(R)}{G_l(R)+iF_l(R)}.
\label{background}
\end{equation}
Notice that the same formulas can be applied to a proton 
case by replacing the spherical Bessel and Neumann functions 
in Eqs. (\ref{FandG}) and (\ref{FandG2}) with 
the corresponding Coulomb wave functions. In this case, the potential 
$V(r)$ is given by, 
\begin{equation}
V(r)=V_0\theta(R-r)+\frac{Ze^2}{r}\theta(r-R), 
\end{equation}
where $Z$ is the charge number of the core nucleus.  

A resonance energy, $E_r$, may be approximately 
defined as the energy at which 
$L_l-S_l$ in the denominator in Eq. (\ref{s-matrix0}) 
vanishes, that is, 
$L_l(E_r)-S_l(E_r)=0$. 
Expanding the quantity $L_l-S_l$ around this energy as, 
\begin{eqnarray}
L_l(E)-S_l(E)&\sim&
L_l(E_r)-S_l(E_r)
-\frac{1}{\gamma^2_l}(E-E_r) 
=-\frac{1}{\gamma^2_l}(E-E_r), 
\label{L-D}
\end{eqnarray}
with 
\begin{equation}
-\frac{1}{\gamma^2_l}\equiv 
\left[\frac{d}{dE}(L_l-S_l)\right]_{E=E_r}, 
\label{gamma0}
\end{equation}
one obtains
\begin{equation}
U_l=e^{2i\delta_l}=\left(1-\frac{i\Gamma_l}{E-E_r+i\frac{\Gamma_l}{2}}\right)
\,e^{2i\phi_l}, 
\end{equation}
where the resonance width $\Gamma_l$ is defined as 
\begin{equation}
\Gamma_l\equiv 2P_l\gamma^2_l.
\label{width0}
\end{equation}
If one neglects the background phase shift, $\phi_l$, 
the $S$-matrix is $U_l=-1$ at $E=E_r$, which is equivalent to $\delta_l=\pi/2$. 
Notice that, because of the presence of the background 
phase shift, in general 
the resonance 
energy, $E_r$, deviates from the energy at which 
the phase shift, $\delta_l$, passes through $\pi/2$. This is the case 
especially for a broad resonance, for which the deviation is significant and 
thus the phase shift does not cross $\pi/2$ at the resonance. 
That is, $\delta_l=\pi/2$ is not equivalent to the resonance 
condition, unless the energy dependence of the background phase shift 
is negligible within the resonance width. 
In that case, one may alternatively locate the resonance 
energy as the energy at which the energy derivative of the phase shift takes 
a maximum as a function of energy \cite{Hagino-Giai04}, since the background 
phase shift is expected to be a slow function of energy. 

A care must be taken for an $s$-wave scattering. In this case, one would 
not expect to have a resonance since there is no centrifugal barrier. 
For a shallow square-well potential, the condition $L_0(E_r)-S_0(E_r)=0$  
may be satisfied at some energy $E_r$. Even in that situation, however, 
the background phase shift is large and the total phase shift, $\delta_0$, 
would not show a clear resonance behavior. Especially, the phase shift would 
not pass through $\pi/2$. For a deeper potential, the phase shift may pass 
through $\pi/2$, but, for $s$-wave scattering, it is always 
downwards in energy \cite{Takeichi}. 
It is then misleading to call it a physical resonance state. 

\begin{figure} [tb]
\begin{center}
\includegraphics[scale=0.6,clip]{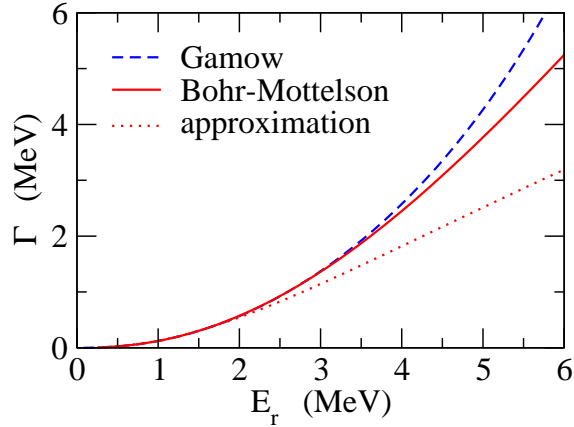}
\end{center}
\caption{
The resonance widths for $d$-wave scattering 
in the mass region of $A\sim$ 30 
as a function of the resonance energy. Here, the 
depth of the square-well potential, $V_0$, is varied 
to obtain different resonance energies, $E_r$.   
The solid line is obtained with the present method, while the 
dashed line is obtained by seeking the Gamow state with a complex 
energy. 
The dotted lines denote 
the results of the approximate formula, Eq. (\ref{gamma_app}), 
for the resonance width for a 
square-well potential. 
}
\end{figure}

Notice that in the present method a pole of the 
$S$-matrix is approximately obtained by neglecting the term $-iP_l$ in 
the denominator in Eq. (\ref{s-matrix0}) to determine the resonance 
energy, $E_r$. This procedure is justified when the resonance 
width is small. 
Figure 1 shows a comparison between 
the resonance widths for $d$-wave scattering 
obtained with two different methods. 
To this end, we 
take the radius parameter in the square-well potential as 
$R=3.95$ fm and vary the depth 
parameter, $V_0$, from $-27$ MeV to $-21$ MeV. 
The mass $\mu$ is taken to be $\mu=30m_N/31$, where $m_N$ is the neutron 
mass, in order to simulate a nucleus in the mass region of $A\sim30$. 
The dashed line in the figure is obtained by seeking the resonance pole without 
an approximation. To this end, 
we impose the 
outgoing boundary 
condition to the radial wave function with a complex energy, 
$E=E_r-i\Gamma/2$. 
The solid line, on the other hand, is obtained with the present 
method by seeking 
the resonance energy 
which satisfies $L_l(E_r)-S_l(E_r)=0$. 
The resonance width is then evaluated according to Eqs. (\ref{gamma0}) 
and (\ref{width0}). 
To this end, 
the logarithmic derivative, $L_l$, is evaluated using the wave function 
for $r<R$ as,  
\begin{equation}
L_l=1+KR\,\frac{j'_l(KR)}{j_l(KR)}, 
\end{equation}
from which the energy derivative of $L_l$ is computed as \cite{bm1}
\begin{equation}
\left.\frac{\partial L_l}{\partial E}\right|_{E=E_r}
=-\frac{\mu R^2}{\hbar^2}
\left[1-\frac{l(l+1)}{K^2R^2}+\frac{S_l(S_l-1)}{K^2R^2}\right]. 
\label{dLdE}
\end{equation}
The energy derivative of $S_l$ in 
Eq. (\ref{gamma0}) is simply computed in a numerical manner. 

In Fig. 1, one can see that the resonance width obtained with 
the present method behaves qualitatively the same as that obtained 
with the complex energy method, although the former width 
tends to be smaller than 
the latter width. See also Appendix A for a comparison 
of the resonance widths obtained with the square-well potential 
to those with a smooth Woods-Saxon potential.

An approximate formula for the resonance width, Eq. (\ref{width0}), 
has been derived in Ref. \cite{bm1} using the 
approximate forms of the spherical Bessel and Neumann functions \cite{Abramowicz}, 
\begin{eqnarray}
j_l(x)&\sim& \frac{x^l}{(2l+1)!!}\left(1-\frac{\frac{1}{2}x^2}{2l+3}+\cdots\right), \\
n_l(x)&\sim& -\frac{(2l-1)!!}{x^{l+1}}
\left(1-\frac{\frac{1}{2}x^2}{1-2l}+\cdots\right), 
\end{eqnarray}
which are valid for $x\ll 1$. 
The resultant formula reads \cite{bm1},
\begin{equation}
\gamma^2_l\sim \frac{\hbar^2}{\mu R^2}\,\frac{2l-1}{2l+1}~~~(l\neq 0,
~kR\ll 1). 
\label{gamma_app}
\end{equation}
This formula has the same dependence on the radius $R$ and the reduced mass $\mu$ 
as the so called Wigner limit for the resonance width \cite{Nunes}. 
However, the underling model wave function is completely different. That is, the Wigner 
limit is obtained with a constant wave function, which would be valid for a non-resonant 
state, while Eq. (\ref{gamma_app}) is obtained with a wave function at a resonance 
condition, which has a large spatial variation. 
Notice that Eq. (\ref{gamma_app}) is not applicable for $l=0$. 
Even though a different form of the approximate 
formula can be derive for $l=0$ \cite{bm1}, we do not discuss it here 
since an $s$-wave hardly forms a resonance in the single-channel problem. 

The resonance widths evaluated with the simple approximate formula, Eq. (\ref{gamma_app}), 
are shown by the dotted line in Fig. 1. 
One can see that the approximate formula indeed works well for small 
values of the resonance energy, i.e., for narrow resonances, while the deviation from the exact 
calculation (the solid line) becomes 
significantly large for broad resonances. 

\begin{figure} [tb]
\begin{center}
\includegraphics[scale=0.6,clip]{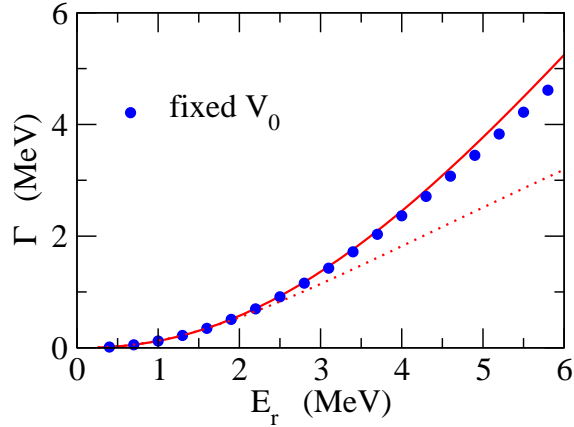}
\end{center}
\caption{
The resonance width obtained by fixing the value of the depth 
parameter of the square-well potential (the filled circles) for the same 
setting as in Fig. 1. The meanings of the solid and the 
dotted lines are the same as in Fig. 1. }
\end{figure}

We notice that the results of the square-well potential are not sensitive 
to the value of the depth parameter, $V_0$. In order to demonstrate this, 
Fig. 2 shows the results obtained with a fixed value of $V_0$, that is, 
$V_0=-27$ MeV (the filled circles). 
Notice that 
Eq. (\ref{dLdE}) and $dS_l/dE$ are functions of energy $E=E_r$ 
for a given value of $V_0$. 
We simply vary the energy, $E_r$, in order to evaluate the resonance 
width, even though the resonance condition, $L_l(E_r)-S_l(E_r)=0$, 
may not be satisfied. 
The solid and the dotted lines are the same as those in 
Fig. 1. One can see that the filled circles closely follow the 
solid line in the region shown in the figure. 
The approximate formula, Eq. (\ref{gamma_app}), is independent of 
the depth parameter. The figure indicates that the resonance width is not 
sensitive to the value of $V_0$ even outside the region in which the 
approximate formula works (that is, $E_r \gtrsim 2$ MeV). 
This is an advantage to use the spherical 
square-well potential, as one does not have 
to consider the consistency between the potential depth and the resonance 
energy. That is, the resonance width can be evaluated easily 
for a given resonance 
energy as an input by taking an arbitrary (but still a reasonable) 
value for the depth parameter. 
Unfortunately, this advantage disappears in multi-channel cases, as 
we discuss in the next section. 

\section{Resonance in a particle-core coupling model}

Let us now extend the formula for the resonance width discussed in 
the previous section to a multi-channel case. To this end, we consider a 
system with a particle coupled to the quadrupole 
motion of a core nucleus, whose potential is 
given by \cite{YH05,Takeichi}, 
\begin{eqnarray}
V(\vec{r})&=&V_0\,\theta
\left(R+R\sum_m\alpha_{2m}Y_{2m}^*(\hat{\vec{r}})-r\right), \\
&\sim& 
V_0\,\theta(R-r) +V_0R\sum_m\alpha_{2m}Y_{2m}^*(\hat{\vec{r}})\,\delta(R-r), 
\end{eqnarray}
where $\alpha_{2m}$ is the collective coordinate for the quadrupole 
motion of the core nucleus. Notice that for simplicity we have expanded 
the potential up to the first order of $\alpha_{2m}$. 

We expand the total wave function as 
\begin{equation}
\Psi_{\alpha_0}(\vec{r})=\sum_\alpha\frac{u_{\alpha\alpha_0}(r)}{r}|\alpha\rangle,
\end{equation}
where $\alpha_0$ denotes the entrance channel (see Eq. (\ref{bc-2}) below) and 
the channel wave function $|\alpha\rangle$ is given by 
\begin{equation}
|\alpha\rangle\equiv|(ljI_c)IM\rangle
=\sum_{m_j,M_c}\langle jm_jI_cM_c|IM\rangle|\phi_{I_cM_c}\rangle
|{\cal Y}_{jlm_j}\rangle. 
\label{chwf}
\end{equation}
Here, $\vec{j}=\vec{l}+\vec{s}$ is the total single-particle angular momentum, 
$I_c$ and $M_c$ are the angular momentum and its $z$-component of the core 
nucleus, respectively. $I$ and $M$ are the total angular momentum of 
the system and its $z$-component, respectively, which are both conserved. 
In Eq. (\ref{chwf}), $|\phi_{I_cM_c}\rangle$ and 
${\cal Y}_{jlm_j}(\hat{\vec{r}})$ are the wave function for the core 
nucleus and the spin-angular wave function for the valence particle, 
respectively. 
When there is no spin-dependent interaction, such as the 
spin-orbit interaction, the spin quantum number is conserved, and one can 
alternatively use the channel wave function given by
\begin{equation}
|\alpha\rangle\equiv|(lI_c)IM\rangle
=\sum_{m_l,M_c}\langle lm_lI_cM_c|IM\rangle|\phi_{I_cM_c}\rangle
|Y_{lm_l}\rangle. 
\label{chwf2}
\end{equation}

We shall use throughout the paper the general 
notation $|\alpha\rangle$ for the channel wave functions, 
which denote either Eq. (\ref{chwf}) or (\ref{chwf2}). 
The coupled-channels equations for the wave functions $u_{\alpha\alpha_0}(r)$ 
then read, 
\begin{eqnarray}
&&\left[-\frac{\hbar^2}{2\mu}\frac{d^2}{dr^2}+V_0\,\theta(R-r)
+\frac{l_\alpha(l_\alpha+1)\hbar^2}{2\mu r^2}+\epsilon_\alpha-E\right]u_{\alpha\alpha_0}(r) \nonumber \\
&&=
-V_0R\,\delta(r-R)\sum_{\alpha'}
\langle\alpha|\sum_m\alpha_{2m}Y_{2m}^*(\hat{\vec{r}})|\alpha'\rangle 
u_{\alpha'\alpha_0}(r) 
\label{cc0}\\
&&\equiv 
-\delta(r-R)\sum_{\alpha'}C_{\alpha\alpha'}
u_{\alpha'\alpha_0}(r), 
\label{cc}
\end{eqnarray}
with 
\begin{equation}
C_{\alpha\alpha'}\equiv
V_0R\,
\langle\alpha|\sum_m\alpha_{2m}Y_{2m}^*(\hat{\vec{r}})|\alpha'\rangle.
\label{coup}
\end{equation} 
Here, $\epsilon_\alpha$ is the excitation energy of the core state, 
$|\phi_{I_cM_c}\rangle$. 
The explicit form of the coupling term, Eq. (\ref{coup}), is given in 
Appendices B and C both for rotational and vibrational cases. 
The solution of the coupled-channels equations for $r<R$ is given as, 
\begin{equation}
u_{\alpha\alpha_0}(r)=A_{\alpha\alpha_0}K_\alpha r j_{l_\alpha}(K_\alpha r)~~~(r < R), 
\label{bc-1}
\end{equation}
with $K_\alpha=\sqrt{\frac{2\mu}{\hbar^2}(E-V_0-\epsilon_\alpha)}$. 
Using the diagonal matrix $\tilde{\vec{F}}(r)$ defined as 
\begin{equation}
\tilde{F}_{\alpha\alpha'}(r)=K_\alpha r j_{l_\alpha}(K_\alpha r)\,\delta_{\alpha,\alpha'}, 
\end{equation}
Eq. (\ref{bc-1}) is also written in a matrix form 
as $\vec{u}(r)=\tilde{\vec{F}}(r)\vec{A}$, where $\vec{u}(r)$ and $\vec{A}$ are 
the matrices 
whose components are given by $u_{\alpha\alpha'}(r)$ and $A_{\alpha\alpha'}$, 
respectively. 
The solution of the coupled-channels equations for $r \geq R$, on the other 
hand, is given by, 
\begin{eqnarray}
u_{\alpha\alpha_0}(r)&=&
k_\alpha r\,h_{l_\alpha}^{(-)}(k_\alpha r)\,\delta_{\alpha,\alpha_0} 
-\sqrt{\frac{k_{\alpha_0}}{k_\alpha}}\,U_{\alpha\alpha_0}
k_\alpha r\,h_{l_\alpha}^{(+)}(k_\alpha r) 
~~~(r \geq R),
\label{bc-2}
\end{eqnarray}
where $U_{\alpha\alpha_0}$ is the $S$-matrix and $k_\alpha$ is defined 
as $k_\alpha=\sqrt{2\mu (E-\epsilon_\alpha)/\hbar^2}$. 
The matching conditions of the wave functions at $r=R$ are given by, 
\begin{eqnarray}
\vec{u}(R_<)&=&\vec{u}(R_>), \label{matching1} \\
-\frac{\hbar^2}{2\mu}(\vec{u}'(R_>)-\vec{u}'(R_<))&=&-\vec{C}\vec{u}(R),
\label{matching2}
\end{eqnarray}
where the prime denotes the radial derivative and $R_>$ and $R_<$ are 
defined as 
$R+\epsilon$ and $R-\epsilon$, respectively, with $\epsilon$ being an 
infinitesimally small number. 
Notice that because of the delta function in Eq. (\ref{cc}) the derivative 
of the wave functions is not continuous at $r=R$. 

When the core nucleus is excited to the channel $\alpha$, the relative energy decreases 
by $\epsilon_\alpha$. Such excitation is kinematically allowed only for $E>\epsilon_\alpha$. 
Those channels which satisfy this condition are referred to as open channels. 
For the channels with $E<\epsilon_\alpha$, the excitations are 
kinematically forbidden and those channels 
are called closed channels. 
Even in this case, the {\it virtual} 
excitations are still possible, which can influence the dynamics 
of the open channels. 
For a closed channel, 
the channel wave number, $k_\alpha$, becomes imaginary, 
$i\kappa_\alpha$, and 
the first term on the right hand side of Eq. (\ref{bc-2}) diverges 
asymptotically. This wave function is thus unphysical. 
Nevertheless, 
the same form of wave functions can be formally employed, if the physical 
$S$-matrix is restricted to the $n\times n$ submatrix of $S$, where $n$ is the number 
of open channels \cite{Taylor}. 

As in the spherical case, 
the wave functions for $r\geq R$, Eq. (\ref{bc-2}), can be solved for the 
matrix $\tilde{U}$, whose components are defined as 
$\tilde{U}_{\alpha\alpha'}=\sqrt{k_{\alpha'}/k_\alpha} U_{\alpha\alpha'}$, and one obtains 
(see Eq. (\ref{s-matrix0})), 
\begin{eqnarray}
\tilde{\vec{U}}&=&
(\vec{G}(R)+i\vec{F}(R))^{-1}
(\vec{L}_>-\vec{S}-i\vec{P})^{-1} 
(\vec{L}_>-\vec{S}+i\vec{P})
(\vec{G}(R)-i\vec{F}(R)), 
\label{Stilde}
\end{eqnarray}
where $\vec{G}(r)$, $\vec{F}(r)$, $\vec{S}$, $\vec{P}$ are diagonal 
matrices whose diagonal components are given 
by $G_{l_\alpha}(r)$, $F_{l_\alpha}(r)$, $S_{l_\alpha}$, 
and $P_{l_\alpha}$, respectively (see Eqs. (\ref{delta}), (\ref{s}), and 
(\ref{FandG})). $\vec{L}_>$ is the logarithmic derivative of the wave 
functions \cite{Johnson78}, 
\begin{equation}
\vec{L}\equiv R\left(\frac{d}{dr}\vec{u}(r)\right)_{r=R}\vec{u}^{-1}(r), 
\end{equation}
evaluated with the wave functions for $r \geq R$. 
Because of the matching conditions, Eqs. (\ref{matching1}) 
and (\ref{matching2}), 
$\vec{L}_>$ is related to $\vec{L}_<$ (that is, the logarithmic derivative 
evaluated with the wave functions for $r < R$) as, 
\begin{equation}
\vec{L}_>=\frac{2\mu R}{\hbar^2}\,\vec{C}+\vec{L}_<.
\end{equation}
Notice that $\vec{L}_<$ is a diagonal matrix, whose components 
are given by
\begin{equation}
(L_<)_{\alpha\alpha'}=\frac{\tilde{F}'_{l_\alpha}(R)}{\tilde{F}_{l_\alpha}(R)}\,\delta_{\alpha,\alpha'},
\end{equation}
with 
$\tilde{F}'_{l}(r)\equiv R \,d(\tilde{F}_{l}(r))/dr$. 

Noticing that 
$\vec{L}_>-\vec{S}+i\vec{P}=\vec{L}_>-\vec{S}-i\vec{P}+2i\vec{P}$ 
and the fact that the matrix 
$(\vec{G}(R)+i\vec{F}(R))^{-1}(\vec{G}(R)-i\vec{F}(R))$ is diagonal with 
the diagonal elements given by Eq. (\ref{background}), Eq. (\ref{Stilde}) can 
be transformed to 
\begin{eqnarray}
\tilde{U}_{\alpha\alpha'}&=&e^{2i\phi_{l_\alpha}}\,\delta_{\alpha,\alpha'} \nonumber \\
&&+2i(G_{l_\alpha}(R)+iF_{l_\alpha}(R))^{-1}
(\vec{L}_>-\vec{S}-i\vec{P})^{-1}_{\alpha\alpha'} 
P_{\alpha'}
(G_{l_\alpha'}(R)-iF_{l_\alpha'}(R)). 
\end{eqnarray}
If we write $G_l(R)-iF_l(R)=\sqrt{G_l(R)^2+F_l(R)^2}\,e^{i\phi_l}$ and 
use Eq. (\ref{s}), the matrix elements of $S$ then read, 
\begin{eqnarray}
U_{\alpha\alpha'}&=&e^{2i\phi_{l_\alpha}}\,\delta_{\alpha,\alpha'} 
+2ie^{i\phi_{l_\alpha}}\sqrt{P_\alpha}
(\vec{L}_>-\vec{S}-i\vec{P})^{-1}_{\alpha\alpha'} 
\sqrt{P_{\alpha'}}e^{i\phi_{l_{\alpha'}}}. 
\label{S-mat}
\end{eqnarray}

We next rewrite the matrix $(\vec{L}_>-\vec{S}-i\vec{P})^{-1}$ as 
\begin{equation}
(\vec{L}_>-\vec{S}-i\vec{P})^{-1}
=
(1-i(\vec{L}_>-\vec{S})^{-1}\vec{P})^{-1}
(\vec{L}_>-\vec{S})^{-1}.
\label{L-D-s}
\end{equation}
Note that the inverse of the matrix $\vec{L}_>-\vec{S}$ can be written as, 
\begin{equation}
(\vec{L}_>-\vec{S})^{-1}
=
\frac{{\rm cof}(\vec{L}_>-\vec{S})}{{\rm det}(\vec{L}_>-\vec{S})},
\end{equation}
where ${\rm det}(\vec{A})$ denotes the determinant of the matrix $\vec{A}$ while 
${\rm cof}(\vec{A})$ is the cofactor matrix transposed, that is, 
\begin{equation}
{\rm det}(A)=\sum_j A_{ij}\,{\rm cof}(A)_{ji},
\label{cofactor}
\end{equation}
for any $i$. 
Suppose that the determinant ${\rm det}(\vec{L}_>-\vec{S})$ is zero at $E=E_r$. 
Then, around this energy the inverse of $\vec{L}_>-\vec{S}$ is approximately given by, 
\begin{equation}
(\vec{L}_>-\vec{S})^{-1}
\sim
-\frac{{\rm cof}(\vec{L}_>-\vec{S})|_{E=E_r}}{d}\,\frac{1}{E-E_r},
\end{equation}
where the coefficient $d$ is defined as (see Eq. (\ref{gamma0})), 
\begin{equation}
d\equiv -\frac{d}{dE}{\rm det}\,(\vec{L}_>-\vec{S})|_{E=E_r}.
\end{equation}
Notice that Eq. (\ref{cofactor}) indicates that when ${\rm det}(A)=0$ the elements of a matrix ${\rm cof}(A)$ is 
given in a separable form as ${\rm cof}(A)_{ij}=c_iy_j$, where $c_i$ is a constant and the vector $\vec{y}$ satisfies 
$\vec{A}\vec{y}=0$. The matrix  $\vec{L}_>-\vec{S}$ is a real symmetric matrix and one can choose the normalization of the 
vector $\vec{y}$ such that  \cite{Taylor}
\begin{equation}
(\vec{L}_>-\vec{S})_{\alpha\alpha'}^{-1}
\sim
-\frac{\gamma_\alpha\gamma_{\alpha'}}{E-E_r},
\label{L-D2}
\end{equation}
with 
\begin{equation}
\gamma^2_\alpha
=
-\frac{[{\rm cof}(\vec{L}_>-\vec{S})|_{E=E_r}]_{\alpha\alpha}}
{\frac{d}{dE}{\rm det}\,(\vec{L}_>-\vec{S})|_{E=E_r}}.
\label{gamma_small}
\end{equation}
Using Eqs. (\ref{S-mat}), (\ref{L-D-s}), and (\ref{L-D2}), 
one finally obtains 
\begin{equation}
U_{\alpha\alpha'}=e^{i\phi_{l_\alpha}}
\left(\delta_{\alpha,\alpha'}
-i\frac{\sqrt{\Gamma_\alpha}\sqrt{\Gamma_{\alpha'}}}
{E-E_r+i\frac{\Gamma_{\rm tot}}{2}}\right)
e^{i\phi_{l_{\alpha'}}},
\end{equation}
with 
\begin{equation}
\Gamma_\alpha=2\gamma^2_\alpha P_\alpha,
\label{gamma}
\end{equation}
and 
\begin{equation}
\Gamma_{\rm tot}=\sum_\alpha \Gamma_\alpha.
\label{gamma_tot}
\end{equation}
This is a well known Breit-Wigner formula for a multi-channel resonance (see, e.g., Refs. \cite{Nunes,LT58,DB10,Weidenmuller67,Hazi79}). 
One may also regard this as a special case of the $R$-matrix formula.  

\begin{figure} [tb]
\begin{center}
\includegraphics[scale=0.6,clip]{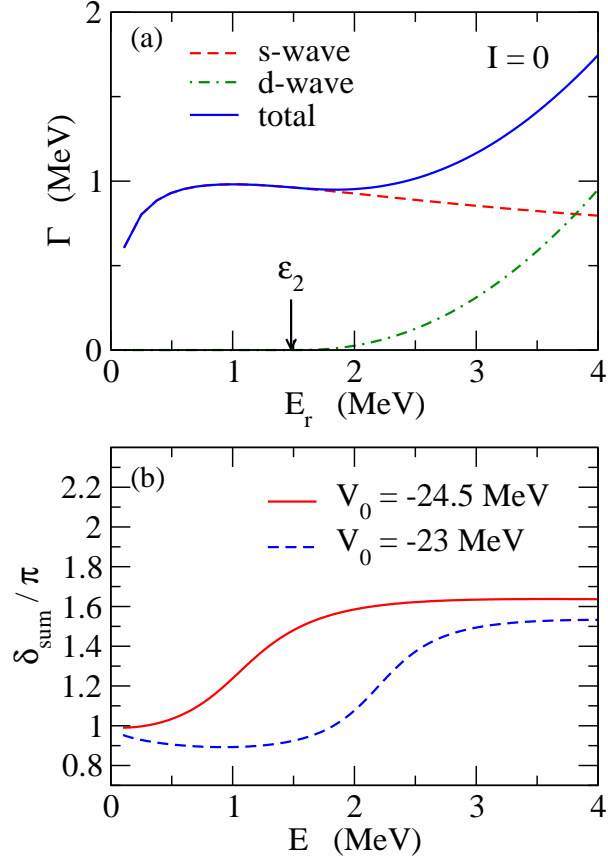}
\end{center}
\caption{
(The upper panel) The resonance widths for a two-channel 
rotational coupling 
with the total angular momentum $I=0$ 
as a function of the resonance energy, $E_r$. 
Here, the depth of the square-well potential, $V_0$, 
is varied to obtain different resonance energies, $E_r$.
The dashed and the dot-dashed lines 
show the partial widths for the 
$|0^+\otimes s\rangle$ and $|[2^+\otimes d]^{(I=0)}\rangle$ channels, 
respectively. 
The threshold energy for the $d$-wave channel ($\epsilon_2=1.48$ MeV) 
is also denoted by the arrow. 
The solid line shows the total width. 
(The lower panel) The eigenphase sums as a function of energy, $E$. 
The solid line is obtained with $V_0=-24.5$ MeV for the depth of the 
square well potential, while the dashed line is obtained with 
$V_0=-23$ MeV.  
The resonance energy and the total width are 
$E_r=0.97$ MeV and $\Gamma_{\rm tot}=0.98$ MeV, respectively, 
for $V_0=-24.5$ MeV, while they are 
$E_r=2.07$ MeV and $\Gamma_{\rm tot}=0.96$ MeV for $V_0=-23$ MeV. 
}
\end{figure}

Notice that a resonance width for a deformed potential, $V(\vec{r})=V_0\theta(R+R\beta_2Y_{20}(\hat{\vec{r}})-r)$ \cite{YH05}, 
can be also evaluated with exactly 
the same formula using the channel wave functions of $|\alpha\rangle=|Y_{lK}\rangle$, $K$ being 
the $z$-component 
of the angular momentum,  and setting  all $\epsilon_\alpha$ to be zero in Eq. (\ref{cc0}). In this case, the matrix elements of 
the coupling potential, Eq. (\ref{coup}), are given as $C_{\alpha\alpha'}=V_0R\,\langle Y_{lK}|Y_{20}|Y_{l'K}\rangle$ \cite{YH05}. 

The upper panel of Fig. 3 shows the result of a two-channel 
calculation. 
For this purpose, we consider the rotational coupling with 
the basis given by Eq. (\ref{chwf2}) with the total angular momentum 
of $I=0$ (see Appendices B and C), that is, a two channel coupling between 
$|0^+\otimes s\rangle$ and $|[2^+\otimes d]^{(I=0)}\rangle$. 
The core nucleus is assumed to be $^{30}$Mg, and we take 
$\epsilon_2$ to be the excitation energy of the first 2$^+$ 
state, that is, 1.48 MeV. The deformation parameter $\beta_2$ is 
estimated to be $\beta_2=0.21$  
by using Eq. (\ref{defpara}) in Appendix B 
with 
the measured $B(E2)$ value, 
$B(E2)\uparrow = 241(31) e^2$fm$^4$ \cite{Niedermaier05} 
(see also Ref. \cite{Mach05}) 
and the radius of 
$R=5.1$ fm 
(that is, $\sqrt{5/3}$ times larger than the radius parameter 
of the square-well potential). 
See Ref. \cite{Nishibata19} for a recent $\beta$-$\gamma$ spectroscopy 
measurement for the $^{31}$Mg nucleus. 
In the figure, the dashed and the dot-dashed curves show the partial 
width for the $s$ and $d$ wave channels, respectively, 
as a function of the resonance energy, $E_r$.
The solid line shows the total width. 
To evaluate these, we vary the depth parameter of the square-well potential 
from $V_0=-24$ MeV to $-19$ MeV. 
For each value of $V_0$, we first 
look for the resonance energy, $E_r$, which satisfies 
${\rm det}(\vec{L}_>(E_r)-S(E_r)=0$.  
We find that this procedure is essential in order to obtain physical values of the 
partial widths, in contrast to the spherical case in which one can 
choose $V_0$ somewhat arbitrarily. After the resonance energy is 
found in this way, 
the partial widths are then evaluated according to Eqs. 
(\ref{gamma}), (\ref{gamma1}), and (\ref{gamma2}). 
For $E<\epsilon_2$, the $d$-wave channel is closed, and the 
partial width for this channel is zero. 
In this case, $S_{l=2}$ is evaluated by changing $k_2$ in the following 
formula \cite{bm1}
\begin{equation}
S_{l=2}=-\frac{18+3k_2^2R^2}{9+3k_2^2R^2+k_2^4R^4}, 
\label{delta2-open}
\end{equation}
to $i\kappa_2$ with $\kappa_2=\sqrt{2\mu |E-\epsilon_2|/\hbar^2}$ 
as 
\begin{equation}
S_{l=2}=-\frac{18-3\kappa_2^2R^2}{9-3\kappa_2^2R^2+\kappa_2^4R^4}. 
\label{delta2-closed}
\end{equation}

\begin{figure} [tb]
\begin{center}
\includegraphics[scale=0.6,clip]{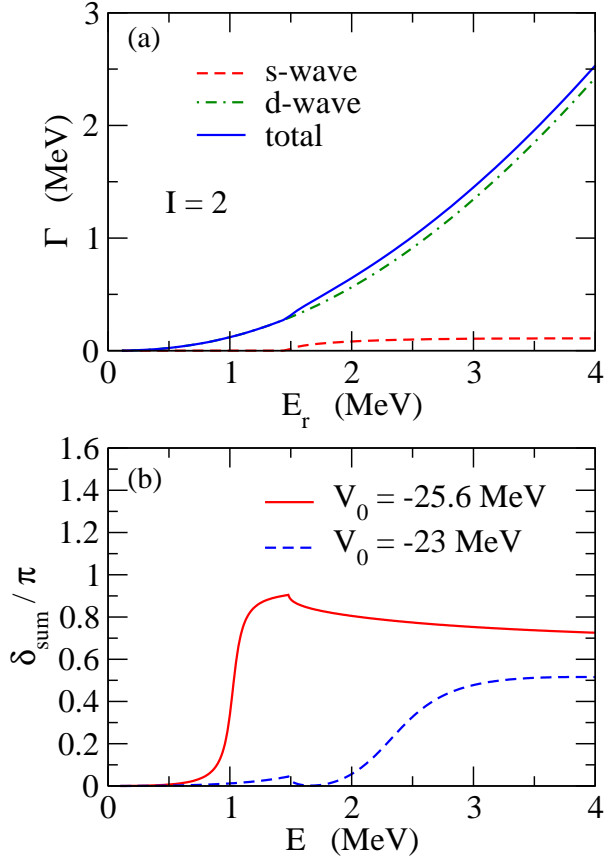}
\end{center}
\caption{
Same as Fig. 3, but for the total angular momentum of $I=2$, with 
channels $|0^+\otimes d\rangle$ and $|2^+\otimes s\rangle$. 
In the lower panel, the eigenphase sums are plotted for 
$V_0=-25.6$ MeV and $-23$ MeV. The resonance energy 
and the width are 
$E_r=0.99$ 
MeV and $\Gamma_{\rm tot}=0.17$ MeV, respectively, for the former potential, 
while they are 
$E_r=2.46$ 
MeV and $\Gamma_{\rm tot}=0.98$ MeV for the latter potential. 
The kinks at the threshold energy for the excited channel ($E=1.48$ MeV) 
are the Wigner cusps. 
}
\end{figure}

One can see that the partial width for the $d$-wave channel becomes 
dominant as the resonance 
energy increases, and thus the resonance state gradually 
changes to a $d$-wave character as a function of energy. 
It is interesting to see that the resonance state exists even when the 
$d$-wave channel is closed at $E_r<\epsilon_2$. 
In order to see this more clearly, the lower panel of Fig. 3 
shows the eigenphase sum, $\delta_{\rm sum}$, defined as 
a sum of eigen phase shifts $\delta_\alpha$, 
that is, 
$\delta_{\rm sum}=\sum_\alpha \delta_\alpha$ 
\cite{Hagino-Giai04,Weidenmuller67,Hazi79} as a function of energy, $E$. 
Here, the eigen phase shifts are 
defined 
as $\lambda_\alpha=e^{2i\delta_\alpha}$, where $\lambda_\alpha$ is an eigenvalue 
of the $S$-matrix. 
The eigenphase sum is a generalization of the phase shift for a 
single-channel case, and shows a resonance behavior with the resonance 
energy $E_r$ and the width $\Gamma_{\rm tot}$ 
\cite{Weidenmuller67,Hazi79}. In the figure, the solid and the dashed 
lines are obtained with $V_0=-24.5$ MeV and $-23$ MeV, respectively. 
Here, we calculate the eigenphase sum 
according to the formula $e^{2i\delta_{\rm sum}}={\rm det}(\vec{U})$ with 
the $S$-matrix computed by Eq. (\ref{S-mat}). 
For the former value of the potential depth, 
the resonance energy and the total width are 
found to be $E_r=0.97$ MeV and $\Gamma_{\rm tot}=0.98$ MeV, respectively, 
while for the 
latter they are 
$E_r=2.07$ MeV and $\Gamma_{\rm tot}=0.96$ MeV. 
One can see that the eigenphase sum increases rapidly around $E\sim E_r$ for each case, 
which is expected as a resonance behavior. 
An interesting thing to see is that the $s$-wave channel shows a resonance 
behavior as a consequence of the coupling to the $d$-wave channel, even 
when the $d$-wave channel is kinetically forbidden (that is, 
a closed channel). 
Another interesting thing is that the square well potential 
with $V_0=-24.5$ MeV does not hold a $d$-wave bound state. 
When 
the energy $E-\epsilon_\alpha$ for a closed channel 
($E-\epsilon_\alpha<0$) coincides with the 
energy of a bound state, 
the phase shift for open channels shows a resonance behavior. This is referred to as 
the Feshbach resonance, and plays an important role in the physics 
of cold atoms \cite{Chin10}. 
Since there is no bound state in the present case, the 
resonance behavior shown by the solid line in the lower panel of 
Fig. 3 should have a different character from a Feshbach resonance. 
Notice that the resonance width is a decreasing function of $E_r$ 
before the $d$-wave channel is open, i.e., for $E_r  < 1.48$ MeV. 
This is partly due to a strong 
energy dependence of $S_2$ given by Eq. (\ref{delta2-closed}), 
which is much stronger than the energy dependence of $S_2$ for 
open channels given by Eq. (\ref{delta2-open}).

The resonance widths for the total 
angular momentum $I=2$ are shown in the upper panel of 
Fig. 4. 
In this case, 
the channels $|0^+\otimes d\rangle$ and $|2^+\otimes s\rangle$ 
are coupled together. One can see that the resonance width is similar 
to the single-channel case shown in the lower panel of Fig. 1. 
That is, the $d$-wave channel dominates the total width with only a small 
contribution from the $s$-wave channel. 
The eigenphase sums for $V_0=-25.6$ MeV and $V_0=-23$ MeV are shown 
in the lower panel of Fig. 4. 
The resonance energy and the total width are 
$E_r=0.99$ 
MeV and $\Gamma_{\rm tot}=0.17$ MeV, respectively, for the former potential, 
while they are 
$E_r=2.46$ 
MeV and $\Gamma_{\rm tot}=0.98$ MeV for the latter potential. 
One can clearly see the expected resonance behavior, especially for the 
former potential with a smaller resonance width. 
In addition, one can also see a kink at the threshold energy for the 
excited channel, that is, at $E=\epsilon_2=1.48$ MeV. 
This is nothing but the Wigner cusp, discussed by Wigner in 1948  
\cite{Wigner} (see also Ref. \cite{MNP07}). 

\begin{figure} [tb]
\begin{center}
\includegraphics[scale=0.6,clip]{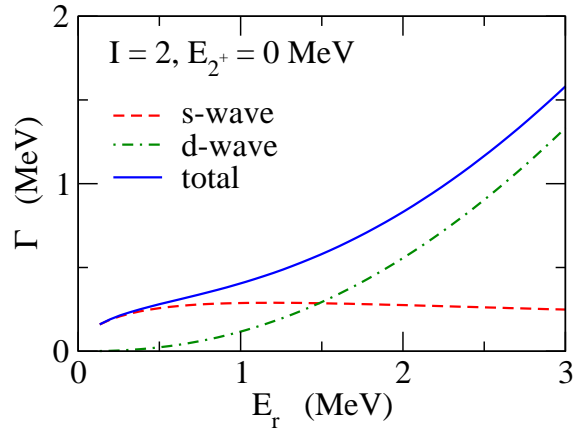}
\end{center}
\caption{
Same as the upper panel of Fig. 4, but obtained by setting the excitation 
energy of the core nucleus to be zero. 
}
\end{figure}

One may have expected that the resonance width for small 
resonance energies 
is determined by the $s$-wave 
due to the absence of the centrifugal barrier while the partial width for 
the $d$-wave channel is largely suppressed. 
This was not seen in Fig. 4 due to the threshold energy, because of which 
the $s$-wave channel contributes only at energies above the threshold.  
In order to gain a deeper insight into 
the role of the $s$-wave channel, we repeat the 
same calculation as in Fig. 4, but by setting the excitation energy 
of the 2$^+$ state of the core nucleus to be zero. The results are shown 
in Fig. 5. As expected, the partial width for the $s$-wave channel is larger 
than that for the $d$-wave channel, at energies below about 0.8 MeV. 
At very low energies, the partial width for the $d$-wave channel 
is largely suppressed due to the finite centrifugal barrier. 
A similar mechanism appears in the two-neutron decay of $^{26}$O, 
for which the resonance energy is extremely small and the decay 
dynamics is largely determined by the 
$s$-wave component in the wave function \cite{Hagino14,Grigorenko13}. 

\section{Summary}

We have derived a compact formula, Eq. (\ref{gamma_small}), 
for partial decay widths 
for a multi-channel resonance state using a deformed square-well 
potential. This was an extension of the formula derived in Ref. \cite{bm1} 
for a single-channel case to a resonance in a particle-core 
coupling model. We have applied the formula to a two-channel 
problem with $s$ and $d$ wave couplings. 
We have shown that,  
even though a pure $s$-wave 
state hardly forms a resonance due to the absence of the centrifugal 
barrier, a resonance may appear as a 
consequence 
of the coupling to the $d$-wave channel. 
This is the case even when the $d$-wave channel is closed and/or 
there is no bound state in the excited channel. 
We have also shown that the $s$-wave component provides a 
dominant contribution at low energies, whenever it is available, while 
the resonance changes to a $d$-wave character as the resonance energy 
increases. 

The resonance formula obtained in this paper is simple and semi-analytic. 
In particular, one does not need to solve the coupled-channels equations 
numerically. The formula is thus useful, at least qualitatively, 
in analyzing experimental data 
e.g., for beta-delayed neutron emissions from neutron-rich nuclei, even 
though several approximations used in this paper may have to be 
carefully examined for quantitative discussions. 
We will present such analysis in a separate paper. 

\section*{Acknowledgments}

We thank M. Ichimura, I. Hamamoto, and T. Shimoda 
for useful discussions. This work was supported in part by JSPS 
KAKENHI  Grant Numbers JP16K05367, JP17J02034, and JP18KK0084.

\appendix

\section{Comparison between a square-well and a Woods-Saxon potentials}

\begin{figure} [tb]
\begin{center}
\includegraphics[scale=0.6,clip]{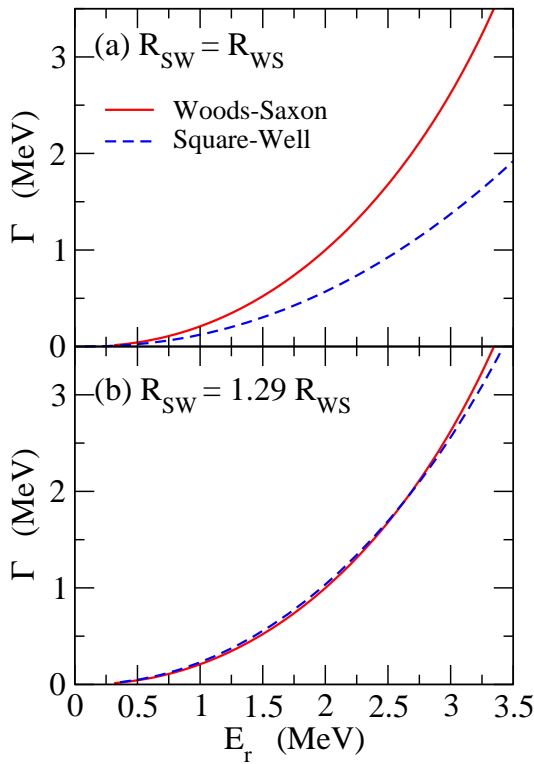}
\end{center}
\caption{
The resonance widths as a function of the resonance energy 
for $d$-wave scattering in the mass region of $A\sim$ 30. 
The solid and the dashed lines show the results of a Woods-Saxon 
and a square-well potentials, respectively. 
In the upper panel, the radius of the 
square-well potential is set to be the same as that of the Woods-Saxon 
potential, while it is increased by a factor of $\sqrt{5/3}$ in the 
lower panel. 
}
\end{figure}

In this Appendix, we compare 
the resonance width 
obtained 
with the spherical square-well (SW) potential 
with that with a smooth Woods-Saxon (WS) 
potential. 
To this end, we seek the $d$-wave Gamow 
state for both the potentials (see the dashed line in Fig. 1).  
For the Woods-Saxon potential, 
we take 
the radius and the diffuseness parameter in the Woods-Saxon as 
$R_{\rm WS}=3.95$ fm and $a_{\rm WS}=0.67$ fm \cite{bm1}, 
respectively, and vary the depth 
parameter, $V^{({\rm WS})}_0$, from $-29$ MeV to $-14$ MeV. 
For the square-well potential, we consider two different choices of the 
radius parameter, $R$. One is to take the same value as in the Woods-Saxon 
potential (the upper panel), and the other is to increase the radius by 
a factor of $\sqrt{5/3}$ (the lower panel) so that the quantity 
\begin{equation}
\langle r^2\rangle = \frac{\int d\vec{r}\,r^2V(r)}{\int d\vec{r}\,V(r)}
\end{equation}
is approximately the same between the Woods-Saxon and the square-well 
potentials. 
For $R=R_{\rm WS}$, we vary the depth parameter $V_0$ 
from $-27$ MeV to $-21$ MeV, while we vary it from $-16$ MeV to $-10.5$ 
MeV for $R=\sqrt{5/3}R_{\rm WS}$. 

The solid and the dashed lines in 
Fig. A1 show the results of the Woods-Saxon and the square-well 
potentials, respectively. 
One can see that the resonance width obtained with 
the square-well potential behaves qualitatively the same as that obtained 
with the Woods-Saxon potential. 
For the increased 
radius parameter (the lower panel), 
the square well potential reproduces even quantitatively 
the result of the Woods-Saxon potential, especially for small values of the 
resonance width. A similar behavior can be found also in absorption 
cross sections, which are relevant to astrophysical fusion reactions 
\cite{Ogura19}. 

\section{Matrix elements of the coupling potential}

In this Appendix, we give explicit forms of the matrix elements of the coupling potential, Eq. (\ref{coup}), with the channel wave functions 
given by Eq. (\ref{chwf}) \cite{HT12}. 
Using Eq. (7.1.6) in Ref. \cite{E57}, 
one obtains 
\begin{eqnarray}
&&V_0R\,\langle(ljI_c)IM|\sum_m \alpha_{2m}Y^*_{2m}
|(l'j'I'_c)IM\rangle \nonumber \\
&=&
V_0R\,(-)^{j'+I_c+I}\left\{
\begin{matrix}
I & I_c & j \\
2 & j' & I'_c \\
\end{matrix}
\right\} \langle {\cal Y}_{jl}||Y_2||{\cal Y}_{j'l'}\rangle \,
\langle \phi_{I_c}||\alpha_2||\phi_{I'_c}\rangle,
\label{angcoup}
\end{eqnarray}
where $\left\{
\begin{matrix}
I & I_c & j \\
2 & j' & I'_c \\
\end{matrix}
\right\}$ denotes the Wigner's  6-$j$ symbol. 
The reduced matrix element $\langle {\cal Y}_{jl}||Y_2||{\cal Y}_{j'l'}\rangle$ is calculated as \cite{E57}, 
\begin{eqnarray}
\langle {\cal Y}_{jl}||Y_2||{\cal Y}_{j'l'}\rangle
&=&\delta_{l+l'+2,{\rm even}}\,(-1)^{\frac{1}{2}+j}\,\frac{\sqrt{5}\,\hat{j}\hat{j}'}{\sqrt{4\pi}}
\,\left(
\begin{matrix}
j & 2& j' \\
1/2 & 0 & -1/2 \\
\end{matrix}
\right), 
\end{eqnarray}
where the bracket denotes the 3-$j$ symbol, and $\hat{j}$ and $\hat{j}'$ are defined as 
$\sqrt{2j+1}$ and $\sqrt{2j'+1}$, respectively. 
In order to evaluate the reduced matrix element, $\langle \phi_{I_c}||\alpha_2||\phi_{I'_c}\rangle$, one needs a specific 
model for  $\phi_{I_c}$, which is given below. 
Notice that the $E2$ transition strength is given by 
\begin{equation}
B(E2: I_c\to I_c')=\frac{1}{2I_c+1}\,|\langle \phi_{I_c}||\hat{T}^{({\rm E2})} ||\phi_{I'_c}\rangle|^2,
\label{be2}
\end{equation}
with the $E2$ operator give by \cite{RS80}
\begin{equation}
\hat{T}_m^{({\rm E2})}=\frac{3e}{4\pi}\,Z_cR^2\alpha_{2m},
\end{equation}
where $Z_c$ is the proton number of the core nucleus. 

\subsection{Vibrational coupling}

We first discuss the vibrational coupling of spherical nuclei. 
In this case, the surface coordinate $\alpha_{2m}$ is regarded as a coordinate for a harmonic oscillator. 
It is related to the phonon creation and annihilation operators as \cite{RS80}
\begin{equation}
\alpha_{2m}=\frac{\beta_2}{\sqrt{5}}\,\left(a_{2m}^\dagger+(-1)^m a_{2-m}\right),
\end{equation}
where $\beta_2/\sqrt{5}$ is the amplitude of the zero-point motion. 
The ground state with $I_c=0$ and the one phonon state with $I_c=2$ are given as, 
\begin{eqnarray}
|\phi_{00}\rangle &=& |0\rangle, \\
|\phi_{2m}\rangle &=& a_{2m}^\dagger|0\rangle,
\end{eqnarray}
respectively, 
where $|0\rangle$ is the vacuum state for the harmonic oscillator. 
From these wave functions, one finds
\begin{equation}
\langle \phi_2||\alpha_2||\phi_0\rangle = \beta_2, 
\end{equation}
and thus \cite{DE01}  
\begin{eqnarray}
&&V_0R\,\langle(ljI_c=2)IM|\sum_m \alpha_{2m}Y^*_{2m}
|(l'j'I'_c=0)IM\rangle 
= V_0R\frac{\beta_2}{\sqrt{4\pi}}\,\left\langle \left.j'\frac{1}{2} 2 0 \right|j \frac{1}{2}\right\rangle.
\label{gs-1st}
\end{eqnarray}
In the vibrational model, there is no coupling from the one phonon state to the same state (that is, the reorientation term). 
From Eq. (\ref{be2}), the value of $\beta_2$ can be estimated from a measured $B(E2)$ strength from the ground state to 
the one phonon state, $B(E2)\uparrow$, as \cite{HT12}
\begin{equation}
\beta_2=\frac{4\pi}{3Z_cR^2}\,\sqrt{\frac{B(E2)\uparrow}{e^2}}.
\label{defpara}
\end{equation}

In a similar way, the wave function for the two phonon states is given as,
\begin{equation}
|\phi_{2{\rm ph}:I_cM_c}\rangle=\frac{1}{\sqrt{2}}\,[a_2^\dagger a_2^\dagger]^{(I_cM_c)}|0\rangle,
\end{equation}   
from which one finds 
\begin{equation}
\langle\phi_{2{\rm ph}:I_c}||\alpha_2||\phi_{1{\rm ph}:I_c=2}\rangle
=\sqrt{\frac{2(2I_c+1)}{5}}\beta_2.
\end{equation}
Similar to the 1 phonon state, 
there is no coupling among the 2 phonon states, as well as the coupling between the 2 phonon states and the ground state. 

\subsection{Rotational coupling}

We next consider the rotational coupling of deformed nuclei. In this case, we first transform the surface coordinate $\alpha_{2m}$ 
to the body-fixed coordinate as 
\begin{equation}
a_{2m}=\sum_{m'}D^2_{m'm}(\varphi_d,\theta_d,\chi_d)\,\alpha_{2m'},
\end{equation}
where $\varphi_d,\theta_d$, and $\chi_d$ are the Euler angles which specify the body-fixed frame, and $D^2_{m'm}$ is the Wigner's 
$D$-function. For axial deformation, only the $m=0$ component in $a_{2m}$ is finite. In this case, one has
\begin{equation}
\alpha_{2m}=\beta_2\sqrt{\frac{4\pi}{5}}\,Y_{2m}(\theta_d,\varphi_d),
\end{equation}
with $a_{20}=\beta_2$ being a deformation parameter. 
For the ground state rotational band in even-even nuclei, the wave function of the rotational states reads 
\begin{equation}
|\phi_{I_cM_c}\rangle = |Y_{I_cM_c}\rangle,
\end{equation}
and one obtains 
\begin{eqnarray}
\langle \phi_{I_c}||\alpha_2||\phi_{I'_c}\rangle
&=&\beta_2\sqrt{\frac{4\pi}{5}}\,\langle Y_{I_c}||Y_2||Y_{I_c'}\rangle 
=(-)^{I_c}\beta_2\hat{I}_c\hat{I}_c'
\left(
\begin{matrix}
I_c & 2& I_c' \\
0 & 0 & 0 \\
\end{matrix}
\right),
\end{eqnarray}
with 
\begin{equation}
\langle Y_{l}||Y_2||Y_{l'}\rangle
=(-1)^{l}\,\frac{\sqrt{5}\,\hat{l}\hat{l}'}{\sqrt{4\pi}}\, 
\left(
\begin{matrix}
l & 2& l' \\
0 & 0 & 0 \\
\end{matrix}\right). 
\label{redmat_l} 
\end{equation}
Notice that the relation between the deformation parameter $\beta_2$ and the $B(E2)$ value is the same as in the vibrational 
case, Eq. (\ref{defpara}). 
The coupling matrix element between the ground state and the first 2$^+$ state is also the same as in the vibrational case and 
is given by Eq. (\ref{gs-1st}).  A difference from the vibrational coupling comes from the reorientation term, that is, the self-coupling 
from the 2$^+$ state to the same state \cite{HT12}. 

\section{Matrix elements with the channel wave functions of 
Eq. (\ref{chwf2})}

In this Appendix, 
we give the matrix elements with the channel wave functions 
given by Eq. (\ref{chwf2}) instead of Eq. (\ref{chwf}). 
The equation which corresponds to Eq. (\ref{angcoup}) in Appendix B 
then reads
\begin{eqnarray}
&&V_0R\,\langle(lI_c)IM|\sum_m \alpha_{2m}Y^*_{2m}
|(l'I'_c)IM\rangle \nonumber \\
&&=
V_0R\,(-)^{l'+I_c+I}\left\{
\begin{matrix}
I & I_c & l \\
2 & l' & I'_c \\
\end{matrix}
\right\} 
\langle Y_{l}||Y_2||Y_{l'}\rangle \,
\langle \phi_{I_c}||\alpha_2||\phi_{I'_c}\rangle.
\end{eqnarray}
The equation that corresponds to Eq. (\ref{gs-1st}) in Appendix B reads
\begin{eqnarray}
&&V_0R\,\langle(lI_c=2)IM|\sum_m \alpha_{2m}Y^*_{2m}
|(l'I'_c=0)IM\rangle 
= V_0R\,\frac{\beta_2}{\sqrt{4\pi}}\,
\langle l'02 0 |l0\rangle, 
\end{eqnarray}
both for the vibrational and for the rotational couplings. 
For the rotational coupling, the reorientation term 
with $l=l'=2$, $I_c=I_c'=2$ and $I=0$ 
is given by \cite{HT12}
\begin{equation}
V_0R\,\langle[d\otimes 2^+]^{(00)}|\sum_m \alpha_{2m}Y^*_{2m}
|[d\otimes 2^+]^{(00)}\rangle 
= V_0R\,\frac{2}{7}\sqrt{\frac{5}{4\pi}}\,\beta_2.
\end{equation}
The reorientation term with $l=l'=0$ vanishes due to the selection 
rule of angular momentum. 

\section{Two-channel coupling}

When only two-channels are involved in the coupling, 
the matrix $\vec{L}_>-\vec{S}$ in Eq. (\ref{L-D2}) is a 2$\times$2 matrix, whose inverse can be 
explicitly written down. 
The matrix $\vec{L}_>-\vec{S}$ has components given by 
\begin{eqnarray}
(\vec{L}_>-\vec{S})_{11}&=&\frac{2\mu}{\hbar^2}RC_{11}+L_1-S_1, \\
(\vec{L}_>-\vec{S})_{22}&=&\frac{2\mu}{\hbar^2}RC_{22}+L_2-S_2, \\
(\vec{L}_>-\vec{S})_{12}&=&(\vec{L}_>-\vec{S})_{21}=\frac{2\mu}{\hbar^2}RC_{12},
\end{eqnarray}
where $S_i~(i=1,2)$ are given by Eq. (\ref{delta}) while $L_i~(i=1,2)$ are given by 
\begin{equation} 
L_i=1+K_iR\,\frac{j'_{l_i}(K_iR)}{j_{l_i}(K_iR)}.
\end{equation}
The determinant of the matrix $\vec{L}_>-\vec{S}$ reads
\begin{equation}
{\rm det}(\vec{L}_>-\vec{S})=(\vec{L}_>-\vec{S})_{11}(\vec{L}_>-\vec{S})_{22}-(\vec{L}_>-\vec{S})_{12}^2.
\end{equation}
Notice that the energy derivative of $L_i$ is given as 
\begin{eqnarray}
\frac{\partial L_i}{\partial E}&=&-\frac{\mu R}{\hbar^2}
\left[1-\frac{l_i(l_i+1)}{K_i^2R^2}+\frac{1}{K_iR}\frac{j'_{l_i}(K_iR)}{j_{l_i}(K_iR)}
+\left(\frac{j'_{l_i}(K_iR)}{j_{l_i}(K_iR)}\right)^2\right].
\end{eqnarray}
In the spherical case, one can use the resonance condition, $L_i=S_i$, to 
obtain the approximate formula for 
$\frac{\partial L_i}{\partial E}$ \cite{bm1}. 
In contrast, in the multi-channel case, the resonance condition is somewhat 
more complicated, that is, ${\rm det}(\vec{L}_>-\vec{S})=0$, and 
a simple approximate formula for 
$\frac{\partial L_i}{\partial E}$ 
cannot be obtained. 

Since the elements of the cofactor matrix of the 2$\times$2 matrix 
$\vec{L}_>-\vec{S}$ are given by
\begin{eqnarray}
{\rm cof}(\vec{L}_>-\vec{S})_{11}&=&(\vec{L}_>-\vec{S})_{22}, \\
{\rm cof}(\vec{L}_>-\vec{S})_{22}&=&(\vec{L}_>-\vec{S})_{11}, 
\end{eqnarray}
the $\gamma^2_i$ in Eq. (\ref{gamma_small}) reads
\begin{eqnarray}
\gamma^2_1
&=&
-\frac{\frac{2\mu}{\hbar^2}RC_{22}+L_2-S_2}
{\frac{d}{dE}{\rm det}\,(\vec{L}_>-\vec{S})|_{E=E_r}}, 
\label{gamma1}
\\
\gamma^2_2
&=&
-\frac{\frac{2\mu}{\hbar^2}RC_{11}+L_1-S_1}
{\frac{d}{dE}{\rm det}\,(\vec{L}_>-\vec{S})|_{E=E_r}},
\label{gamma2}
\end{eqnarray}
where the numerators are evaluated at $E=E_r$. 
The partial and the total widths are then evaluated according 
to Eqs. (\ref{gamma}) and (\ref{gamma_tot}), respectively.  

In the no-coupling limit, the coupling matrix $\vec{C}$ vanishes, 
and thus the determinant of $\vec{L}_>-\vec{S}$ becomes
\begin{equation}
{\rm det}(\vec{L}_>-\vec{S})=(L_1-S_1)(L_2-S_2). 
\end{equation}
Suppose that $L_1-S_1=0$ at $E=E_r$. Then, the energy derivative 
of the determinant reads
\begin{equation}
\frac{d}{dE}\left.{\rm det}(\vec{L}_>-\vec{S})\right|_{E=E_r}=
-\frac{1}{\gamma^2_{l_1}}(L_2-S_2), 
\end{equation}
where $\gamma^2_{l_1}$ is given by Eq. (\ref{gamma0}). 
This leads to $\gamma^2_1=\gamma^2_{l_1}$ and $\gamma^2_2=0$, which is consistent 
with the single-channel case discussed in Sec. II.


\begin{thebibliography}{99}
\bibitem{phase}We follow Ref. \cite{Abramowicz} to 
use the phase convention for the spherical Neumann 
function, with which the asymptotic form is given by 
$n_l(kr)\to -\cos(kr-l\pi/2)~~~(r\to\infty)$. 
Notice that this is a different sign convention from that 
used in Ref. \cite{bm1}.
\end{thebibliography}
\end{document}